\documentclass[a4paper,10pt,aps,pra,nofootinbib]{revtex4-1}
\usepackage{amsmath,amsthm}
\usepackage{comment,amssymb,latexsym,upref,enumerate,fouridx}

\usepackage{finalT}

\newcommand{\Ahu}{\underline{\hat{A}}}
\newcommand{\Bhu}{\underline{\hat{B}}}
\newcommand{\thetau}{\underline{\theta}}

\numberwithin{equation}{section}

\DeclareMathAlphabet{\mathpzc}{OT1}{pzc}{m}{it}

\DeclareMathOperator{\Div}{div}

\newcommand{\id}{\mathord{{\mathrm 1}\kern-0.27em{\mathrm I}}\kern0.35em}

\newcommand{\del}[1]{\partial_{#1}}

\newcommand{\Ord}[2]{\text{O}(#1^{#2})}

\newcommand{\oset}[2]{\overset{#1}{#2}{}}

\newcommand{\AND}{{\quad\text{and}\quad}}

\newcommand{\Fc}{\mathcal{F}{}}

\newcommand{\Rc}{\mathcal{R}{}}

\newcommand{\Ut}{\tilde{U}{}}

\newcommand{\Vt}{\tilde{V}{}}

\newcommand{\xt}{\tilde{x}{}}

\newcommand{\yt}{\tilde{y}{}}

\newcommand{\thetat}{\tilde{\theta}{}}

\newcommand{\psit}{\tilde{\psi}{}}

\newcommand{\Ah}{\hat{A}{}}

\newcommand{\Bh}{\hat{B}{}}

\newcommand{\Hh}{\hat{H}{}}

\newcommand{\Th}{\hat{T}{}}

\newcommand{\vv}{\mathbf{v}{}}

\newcommand{\Rbb}{\mathbb{R}{}}

\newcommand{\ep}{\epsilon}

\begin{document}

\title{Classical-quantum limits}
\author{Todd A. Oliynyk}
\affiliation{School of Mathematical Sciences\\
Monash University, VIC 3800\\
Australia}
\email{todd.oliynyk@monash.edu}

\begin{abstract}
\noindent  We introduce a new approach to analyzing the interaction between classical and quantum
systems that is based on a limiting procedure applied to multi-particle Schr\"{o}dinger equations. The
limit equations obtained by this procedure, which we refer to as the classical-quantum limit, govern the interaction between classical and quantum systems, and
they possess many desirable properties that are inherited
in the limit from the multi-particle quantum system.
As an application, we use the classical-quantum limit equations to identify the source of the non-local
signalling that is known to occur in the classical-quantum hybrid scheme of Hall and Reginatto. We also
derive the first order correction to the classical-quantum limit equation to obtain a fully consistent first order approximation
to the Schr\"{o}dinger equation that should be accurate for modeling the interaction between
particles of disparate mass in the regime where the particles with the larger masses are effectively classical.
\end{abstract}

\maketitle

\sect{intro}{Introduction}

Consistent schemes for the interaction of classical and quantum systems are of great interest from both theoretical and practical points of view. For example, in the Copenhagen interpretation of quantum mechanics,
measuring devices are
treated as classical objects that interact with quantum systems. Consequently, from a foundational perspective, having a consistent and effective method for analyzing the interaction between classical
and quantum systems is highly desirable. From the perspective of computational chemistry, the ability to treat
heavy particles classically and light particles quantum mechanically results in a significant reduction in
computational costs over fully quantum mechanical treatments. For systems with large particle numbers, this separation is essential to making computations feasible.

There has been much work done on the subject of the interaction of classical and quantum systems, and
various schemes have been developed to investigate these types of interactions. In the literature
these are known as classical-quantum hybrid schemes, and a wide range of distinct schemes have been
developed for various purposes including modeling of chemical reactions,
analyzing decoherence, and studying measurement theory \cite{Anderson:1995,BoucherTraschen:1988,Bousquet_et_al:2011,DiosiHalliwell:1998,
Diosi_et_al,Fratino_et_al:2014,Hall:2008,HallReginatto:2005,Horenko_et_al:2004,Kapral:1999,Kisil:2005,
Lampo_et_al:2014,Meier:1993,PeresTerno:2001,PrezhdoBrooksby:2001,PrezhdoKisil:1997,ReginattoHall:2009}. Additionally, a number of studies of the
general structure and properties of classical-quantum hybrid schemes have been carried out \cite{Agostini_et_al:2007,Elze:2012a,Elze:2012b,Elze:2013,CaroSalcedo:1999,Sahoo:2003,Salcedo:1996,Salcedo:2012}.

Although many classical-quantum hybrid schemes have been introduced, none are regarded as being the definitive solution to the problem of describing the interaction between
classical and quantum systems. In this article, we introduce a new method to address this problem that is based on taking
a singular limit of a fully quantum system in complete analogy with the more well-known classical limit.
More specifically, we introduce a small dimensionless
parameter $\ep$ into a multi-particle Schr\"{o}dinger equation in such a way that the parameter only appears in a subset of the single particle Hamiltonian operators.
Using a non-local polar representation for the wave function,
we show that the multi-particle Schr\"{o}dinger equation, when expressed
in terms of our non-local polar decomposition, admits a formal $\ep \searrow 0$ limit. The resulting limit equations
define our notion of a classical-quantum limit, which as far as we are aware is new.
The derivation of these equations is carried out in Section \ref{cqlim}, and we also justify our interpretation
of these limit equations as describing the interaction between a classical and quantum system in that section.

The limit equations derived here define a completely natural classical-quantum hybrid scheme that has many attractive properties.
We single out two properties.
First,  since our limit equations arise from the taking the singular limit of a fully quantum mechanical system, the desirable properties that hold true for
the quantum system remain true for limit equations for any quantities with a well defined limit.
In particular,
this implies that the limit equations that arise from a non-interacting quantum system will satisfy the non-signalling property
that is known to be satisfied by non-interacting quantum systems. This fact forms the basis of our first application where in Section \ref{app1} we use our
classical-quantum limit equations to clarify the origin of the non-local signalling \cite{Hall_et_al:2012} that was observed to occur
in the classical-quantum hybrid scheme of
Hall and Reginatto \cite{Hall:2008,HallReginatto:2005,ReginattoHall:2009}.
Second, again due to the limiting
procedure, it can be
expected that solutions of the limit equations will approximate corresponding fully quantum
mechanical solutions, at least on suitable time scales as is known to be true in the classical limit \cite{Bardos_et_al:2015}.
From this perspective, our limit equations capture the interactions between
classical and quantum systems in a manner that is fully consistent with quantum theory.
Moreover, we can view our limit equations as the zeroth order equations
satisfied by an $\ep$ expansion of $\ep$-dependent solutions to the fully quantum mechanical equations.
Using a standard perturbation expansion method, we derive in Section \ref{app2} the equations satisfied by the first order
expansion term. In this way, it is possible to derive
a fully consistent perturbative approximation scheme for multi-particle quantum systems where the base approximation consists
of an interacting classical-quantum system. We expect that this scheme will be useful in computational chemistry
for modelling the interaction of heavy particles that are treated in the first approximation classically with light particles
that are treated quantum mechanically.

\sect{cqlim}{A classical-quantum limit}

Our derivation of the classical-quantum limit begins with a 2-particle Schr\"{o}dinger equation
of the form
\leqn{SchrodUS}{
i\hbar \del{t}\psi = -\frac{\hbar^2}{2 M_1}\del{x}^2\psi + U\psi
-\frac{\hbar^2}{2 M_2}\del{y}^2\psi + V\psi,
}
where $U = U(x)$ and  $V=V(x,y)$
are time-independent potentials and we are using $\del{x}$ and $\del{y}$ to denote partial derivatives, i.e.
\eqn{pdiffs}{
\del{x}\phi = \frac{\del{}\phi}{\del{}x} \AND \del{y}\phi = \frac{\del{}\phi}{\del{}y}
}
for any function $\phi=\phi(x,y)$.

\begin{rem}
In the following, the partial derivatives will act on functions that directly follow them. For example,
\begin{equation*}
\del{x}\phi\psi = \frac{\del{}\phi}{\del{}x}\,\psi.
\end{equation*}
If we want the partial derivative to act on both functions, then we will employ brackets and write
\begin{equation*}
\del{x}(\phi\psi) =  \frac{\del{}(\phi\psi)}{\del{}x}.
\end{equation*}
\end{rem}

Next, we introduce time and length scales given by $T$, $L_1$ and $L_2$, and define the dimensionless variables
\gath{dimvars}{
\tilde{t} = \frac{t}{T},  \quad
\xt = \frac{x}{L_1}, \quad
\yt = \frac{y}{L_2}, \\
\psit(\tilde{t},\xt,\yt) = \psi(T\tilde{t},L_1\xt,L_2 \yt),\\
\Ut(\xt) = \frac{T M_2 }{\hbar M_1 }U(L_1\xt)
\intertext{and}
\Vt(\xt,\yt) = \frac{T}{\hbar}V(L_1\xt,L_2\yt).
}
Writing the Schr\"{o}dinger equation \eqref{SchrodUS} in terms of these dimensionless variables, we see that $\psit$
satisfies
\leqn{SchrodS}{
i\del{\tilde{t}}\psit = -\frac{\hbar T}{2 M_1 L_1^2}\del{\xt}^2\psit + \frac{M_1}{M_2}\Ut\psit
-\frac{\hbar T}{2 M_2 L_2^2}\del{\yt}^2\psit + \Vt\psit.
}

We introduce a small dimensionless parameter $\ep$ via
\leqn{epdef}{
\ep = \frac{M_2}{M_1}.
}
Defining the dimensionless masses
\eqn{mtdef}{
m_1 = \frac{M_2 L_1^2}{\hbar T}  \AND m_2 = \frac{M_2 L_2^2}{\hbar T},
}
and dropping the tildes, we can write the dimensionless two-particle Schr\"{o}dinger equation \eqref{SchrodS} as
\leqn{Schrod}{
i\del{t}\psi^\ep = -\frac{\ep}{2m_1}\del{x}^2\psi^\ep + \frac{1}{\ep}U\psi^\ep
-\frac{1}{2 m_2}\del{y}^2\psi^\ep + V\psi^\ep,
}
where we have written $\psi=\psi^\ep$ to emphasize that the solution depends on the small parameter $\ep$.
We remark that we restrict ourselves to a pair of 1-dimensional particles and time-independent potentials for simplicity.
All of our results generalize to higher dimensions, more particles and time-dependent potentials in
an obvious manner.

The equation \eqref{Schrod}
governs the evolution of a pair of 1-dimensional quantum particles of (dimensionless) mass $m_1/\ep$ and $m_2$, respectively.
Assuming the length scales $L_1$ and $L_2$ are comparable, the ratio $(m_1/\ep)/m_2 \approx 1/\ep$ is unbounded as $\ep \searrow 0$ showing
that the limit $\ep \searrow 0$ is an extreme mass ratio limit that represents the interaction of a heavy particle with a light one. Furthermore, due to the
positioning of the parameter $\ep$, it is clear that the $\ep \searrow 0$ limit acts like a classical
limit, but only for the first system with the configuration coordinate $x$.
From these considerations,
we expect that a consistent
$\ep \searrow 0$ limit represents the interaction of a classical particle with a quantum one. We justify
this expectation in the remainder of this article.

\subsect{modpolar}{A non-local polar representation}

The classical-quantum limit considered here is based on a non-local polar representation
for the wave function where the non-locality is introduced by way of the averaging operator
\leqn{Ah1}{
\Ah(\phi)(x,y) := \frac{1}{(b-a)}\int_{a}^b \phi(x,z)\, dz 
}
with $a,b\in \Rbb$ fixed constants satisfying $a<b$.
We observe that $\Ah$ satisfies
\leqn{Ah2}{
\Ah^2 = \Ah,
}
and hence, defines a projection operator. It is also easy to see that $\Ah$ satisfies the properties
\leqn{Ah3}{
\del{x}\Ah = \Ah \del{x} \AND \del{y}\Ah = 0, 
}
and that
\leqn{Ah4}{
\Ah\phi = \phi \AND \Ah(\phi\xi)= \phi \Ah\xi \qquad \text{for $\phi=\phi(x)$ and $\xi=\xi(x,y)$}.
}

\begin{rem} \label{Arem}
The properties \eqref{Ah2}-\eqref{Ah4} are the essential properties of the operator $\Ah$ that are
needed to define our classical-quantum limit. Given any operator $\Ah$ satisfying these properties,
all of the arguments used in this article remain valid, and hence, there exists a classical-quantum limit
associated with every operator $\Ah$ satisfying \eqref{Ah2}-\eqref{Ah4}, and in particular, for each choice of
$a,b\in \Rbb$, $a<b$. These different classical-quantum
limits are all related via a bijective transformation, and consequently, different choices of $\Ah$
lead to different representations of the \emph{same classical-quantum limit}; see Appendix \ref{crep} for
the details. Thus there is no physical content in the particular choice of operator $\Ah$
satisfying  \eqref{Ah2}-\eqref{Ah4}, and throughout this article, we will employ the operator defined by
\eqref{Ah1} for a fixed choice of constants $a,b\in \Rbb$, $a<b$.
\end{rem}

Exponentiating $\Ah$ yields
\leqn{Ahexp}{
\exp(\tau \Ah) = e^\tau \Ah + \Bh
}
where
\eqn{Bh1}{
\Bh := \id - \Ah
}
is the complementary projection operator, which, by \eqref{Ah2}, \eqref{Ah3} and \eqref{Ah4}, satisfies
\lgath{Bh2}{
\Bh^2 = \Bh, \quad \Bh\Ah = \Ah \Bh = 0, \quad
\Ah + \Bh =\id, \label{Bh2.1}\\
 \del{x}\Bh = \Bh \del{x}, \quad \del{y}\Bh =\del{y}, 
}
and
\leqn{Bh3}{
\Bh\phi = 0 \AND \Bh(\phi\xi)=\phi\Bh\xi \quad \text{for $\phi=\phi(x)$ and $\xi=\xi(x,y)$.}
}

Letting
\leqn{Rthetadef}{
\psi^\ep(x,y) = e^{i\Theta^\ep(x,y)}R^\ep(x,y) \qquad (R^\ep(x,y):=|\psi^\ep(x,y)|)
}
denote the standard polar representation, we introduce a non-local phase
variable $\theta^\ep$ via the definition
\leqn{thetadef}{
\theta^\ep := \exp(\ln(\ep)\Ah)\Theta^\ep \oset{\eqref{Ahexp}}{=} \ep \Ah \Theta^\ep + \Bh \Theta^\ep,
}
or equivalently
\leqn{thetadef1}{
\Theta^\ep = \exp(-\ln(\ep)\Ah)\theta^\ep \oset{\eqref{Ahexp}}{=} \frac{1}{\ep} \Ah \theta^\ep + \Bh \theta^\ep.
}
Substituting \eqref{Rthetadef} into \eqref{Schrod}, we find using \eqref{thetadef1}
that the pair $(R^\ep,\theta^\ep)$ satisfies
\lalin{Schrod1}{
\del{t}R^\ep &= -\frac{\ep}{m_1}\del{x}\bigl(\exp(-\ln(\ep)\Ah)\theta^\ep \bigr)\del{x}R^\ep -
 \frac{\ep}{2m_1}\del{x}^2\bigl(\exp(-\ln(\ep)\Ah)\theta^\ep \bigr)R^\ep \notag \\
& \hspace{2.0cm} -\frac{1}{m_2}\del{y}\bigl(\exp(-\ln(\ep)\Ah)\theta^\ep \bigr)\del{y}R^\ep - \frac{1}{2 m_2}\del{y}^2\bigl(\exp(-\ln(\ep)\Ah)\theta^\ep \bigr)R^\ep, \label{Schrod1.1}\\
\del{t}\theta^\ep &= \exp(\ln(\ep)\Ah)\biggr\{-\frac{\ep}{2 m_1}\bigl[\del{x}\bigl(\exp(-\ln(\ep)\Ah)\theta^\ep\bigr)\bigr]^2
-\frac{1}{\ep}U + \frac{\ep}{2 m_1} \frac{\del{x}^2 R^\ep}{R^\ep} \notag \\
& \hspace{3.5cm}  -\frac{1}{2 m_2}\bigl[\del{y}\bigl(\exp(-\ln(\ep)\Ah)\theta^\ep\bigr)\bigr]^2
 + \frac{1}{2 m_2} \frac{\del{y}^2 R^\ep}{R^\ep} - V \label{Schrod1.2}
\biggl\},
}
which is completely equivalent to the 2-particle Schr\"{o}dinger equation \eqref{Schrod} for any $\ep > 0$.
We claim that this system has a regular $\ep\searrow 0$ limit. To see this, we observe, from
repeated use of the relations \eqref{Ah2}-\eqref{Bh3}, that the following terms that appear to be singular
are in fact regular:
\lalin{obs1}{
\del{y}\bigl(\exp(-\ln(\ep)\Ah)\theta^\ep\bigr) & = \del{y}\biggl(\frac{1}{\ep} \Ah \theta^\ep +
\Bh \theta^\ep\biggr) = \del{y}\Bh\theta^\ep
=\del{y}\theta^\ep, \label{obs1.1}\\
\exp(\ln(\ep)\Ah)\frac{1}{\ep}U &= \frac{1}{\ep}(\ep\Ah +\Bh)\Ah U = U, \label{obs1.2}
}
and
\lalin{obs2}{
\exp(\ln(\ep)\Ah)\ep\bigl[\del{x}\bigl(\exp(-\ln(\ep)\Ah)\theta^\ep\bigr)\bigr]^2 &=
\ep\bigl(\ep \Ah + \Bh)\biggl(\frac{1}{\ep^2}\bigl[\del{x}\Ah\theta^\ep\bigr]^2
+\frac{2}{\ep}\del{x}\Ah \theta^\ep\del{x}\Bh\theta^\ep +\bigl[\del{x}\Bh\theta^\ep\bigr]^2 \biggr)  \notag \\
& = \bigl(\del{x}\Ah\theta^\ep\bigr)^2
 +\ep^2 \Ah\bigl(\del{x}\Bh \theta^\ep\bigr)^2
+
2\del{x}\Ah\theta^\ep\del{x}\Bh\theta +\ep\Bh\bigl(\del{x}\Bh \theta^\ep\bigr)^2 . \label{obs2.1}
}

Using \eqref{obs1.1}, \eqref{obs1.2} and \eqref{obs2.1} along with the relations \eqref{Ah2}-\eqref{Bh3}, it is not difficult
to verify that \eqref{Schrod1.1}-\eqref{Schrod1.2} can be written as
\lalin{Schrod2}{
\del{t}R^\ep &= -\frac{1}{m_1}\bigl(\del{x}\Ah\theta^\ep + \ep\del{x}\Bh\theta^\ep\bigr)\del{x}R^\ep -
 \frac{1}{2m_1}\bigl(\del{x}^2\Ah\theta^\ep + \ep \del{x}^2\Bh\theta^\ep \bigr)R^\ep -\frac{1}{m_2}\del{y}\theta^\ep \del{y}R^\ep - \frac{1}{2 m_2}\del{y}^2\theta R^\ep, \label{Schrod2.1}\\
\del{t}\theta^\ep &= -\frac{1}{2 m_1}\Bigl[\bigl(\del{x}\Ah\theta^\ep\bigr)^2
 +\ep^2 \Ah\bigl(\del{x}\Bh \theta^\ep\bigr)^2 +
2\del{x}\Ah\theta^\ep\del{x}\Bh\theta^\ep +\ep\Bh\bigl(\del{x}\Bh \theta^\ep\bigr)^2\Bigr]
-U \notag \\
& \hspace{2.5cm} + (\ep \Ah + \Bh)\biggl[ \frac{\ep}{2 m_1} \frac{\del{x}^2 R^\ep}{R^\ep}
  -\frac{1}{2 m_2}\bigr(\del{y}\theta^\ep\bigr)^2
 + \frac{1}{2 m_2} \frac{\del{y}^2 R^\ep}{R^\ep} - V
\biggl]. \label{Schrod2.2}
}
Setting
\leqn{thetaAB}{
\theta^\ep_A := \Ah\theta^\ep \AND \theta^\ep_B := \Bh\theta^\ep,
}
we see after applying the projection operators $\Ah$ and $\Bh$ to \eqref{Schrod2.2}, and again using the relations \eqref{Ah2}-\eqref{Bh3}
that \eqref{Schrod2.1}-\eqref{Schrod2.2} are equivalent to
\lalin{Schrod2a}{
\del{t}R^\ep &= -\frac{1}{m_1}\bigl(\del{x}\theta^\ep_A + \ep\del{x}\theta^\ep_B\bigr)\del{x}R^\ep -
 \frac{1}{2m_1}\bigl(\del{x}^2\theta^\ep_A + \ep \del{x}^2\theta^\ep_B \bigr)R^\ep -\frac{1}{m_2}\del{y}\theta^\ep_B
 \del{y}R^\ep - \frac{1}{2 m_2}\del{y}^2\theta_B R^\ep, \label{Schrod2a.1}\\
\del{t}\theta^\ep_A &= -\frac{1}{2 m_1}\Bigl[\bigl(\del{x}\theta^\ep_A\bigr)^2
 +\ep^2 \Ah\bigl(\del{x}\theta^\ep_B\bigr)^2\Bigr]
-U + \ep \Ah \biggl[ \frac{\ep}{2 m_1} \frac{\del{x}^2 R^\ep}{R^\ep}
  -\frac{1}{2 m_2}\bigr(\del{y}\Bh\theta^\ep\bigr)^2
 + \frac{1}{2 m_2} \frac{\del{y}^2 R^\ep}{R^\ep} - V
\biggl], \label{Schrod2a.2} \\
\del{t}\theta^\ep_B &= \Bh\biggl[-\frac{1}{2 m_1}\Bigl(
2\bigl(\del{x}\theta^\ep_A\del{x}\theta^\ep_B\bigr) +\ep\bigl(\del{x}\theta^\ep_B\bigr)^2\Bigr)
 +\  \frac{\ep}{2 m_1} \frac{\del{x}^2 R^\ep}{R^\ep}
  -\frac{1}{2 m_2}\bigr(\del{y}\theta^\ep_B\bigr)^2
 + \frac{1}{2 m_2} \frac{\del{y}^2 R^\ep}{R^\ep} - V
\biggl]. \label{Schrod2a.3}
}
It is worth noting that we can treat $\theta^\ep_A=\theta^\ep_A(t,x)$ and $\theta^\ep_B=\theta^\ep_B(t,x,y)$ as independent variables provided that the initial data for $\theta_B$ at $t=0$ is chosen to
satisfy the constraint $\Ah \theta^\ep_B|_{t=0} = 0$. This is possible because it follows easily from \eqref{Schrod2a.3}
that the constraint $\Ah \theta^\ep_B = 0$ propagates, i.e. satisfies $\del{t}\Ah \theta^\ep_B =0$, from which it follows that the
function $\theta^\ep$ defined by $\theta^\ep(t,x,y) = \theta^\ep_A(t,x) + \theta^\ep_B(t,x,y)$ verifies
 the relations $\theta^\ep_A = \Ah \theta^\ep$ and $\theta^\ep_B = \Bh \theta^\ep$ in accordance with  \eqref{thetaAB}.

To summarize, the two systems \eqref{Schrod2.1}-\eqref{Schrod2.2} and \eqref{Schrod2a.1}-\eqref{Schrod2a.3}
represent equivalent formulations of the 2-particle Schr\"{o}dinger equation \eqref{Schrod} under
the transformation
\leqn{Schrodtrans1}{
\psi^\ep= e^{i\bigl(\frac{1}{\ep}\theta^\ep_A+ \theta^\ep_B\bigr)}R^\ep
}
where
\leqn{Schrodtrans2}{
\theta^\ep = \theta^\ep_A + \theta^\ep_B,
}
and $\theta^\ep_A$ and $\theta^\ep_B$ are given in terms of $\theta^\ep$ by \eqref{thetaAB}. Moreover, it is clear from \eqref{Schrodtrans1}, that if
$\chi_\Omega$ is the characteristic function of a subset $\Omega \subset \Rbb^2$, i.e. $\chi_\Omega(x,y)=1$ if $(x,y)\in \Omega$ and $0$ otherwise,
then the action of the operator $\chi_\Omega(\hat{x},\hat{y})$ on the wave function $\psi^\ep$ is equivalent to the following action on
the pair $(\theta^\ep,R^\ep)$:
\eqn{measure1}{
\chi_\Omega(\hat{x},\hat{y})\bigl(\theta^\ep,R^\ep\bigr)(x,y) = (\theta^\ep(x,y),\chi_\Omega(x,y)R^\ep(x,y)).
}
From this, we conclude that position measurements are represented by $\ep$-independent maps in our non-local polar representation.

\subsect{limiteqns}{The limit equations}

Taking the formal $\ep\searrow 0$ limit
\eqref{Schrod2.1}-\eqref{Schrod2.2} and
\eqref{Schrod2a.1}-\eqref{Schrod2a.3}, we obtain the two equivalent forms of
the
limit equations given by
\lalin{Schrodlim0}{
\del{t}R &= -\frac{1}{m_1}\del{x}(\Ah\theta)\del{x}R -
 \frac{1}{2m_1}\del{x}^2(\Ah\theta)R -\frac{1}{m_2}\del{y}\theta \del{y}R - \frac{1}{2 m_2}\del{y}^2\theta R, \label{Schrodlim0.1}\\
\del{t}\theta &= -\frac{1}{2 m_1}\Bigl[\bigl(\del{x}\Ah\theta\bigr)^2 + 2\del{x}\Ah\theta\del{x}\Bh\theta\Bigr]
-U  +\Bh\biggl[
  -\frac{1}{2 m_2}\bigr(\del{y}\theta\bigr)^2
 + \frac{1}{2 m_2} \frac{\del{y}^2 R}{R} - V
\biggl], \label{Schrodlim0.2}
}
and
\lalin{Schrodlim1}{
\del{t}R &= -\frac{1}{m_1}\del{x}\theta_A \del{x}R -
 \frac{1}{2m_1}\del{x}^2\theta_A R -\frac{1}{m_2}\del{y}\theta_B \del{y}R - \frac{1}{2 m_2}\del{y}^2\theta_B R, \label{Schrodlim1.1}\\
\del{t}\theta_A &= -\frac{1}{2 m_1}\bigl(\del{x}\theta_A\bigr)^2
-U, \label{Schrodlim1.2} \\
\del{t}\theta_B &= -\frac{1}{m_1}
\del{x}\theta_A\del{x}\theta_B
+ \Bh\biggl[
  -\frac{1}{2 m_2}\bigr(\del{y}\theta_B\bigr)^2
 + \frac{1}{2 m_2} \frac{\del{y}^2 R}{R} - V
\biggl], \label{Schrodlim1.3}
}
respectively. In the second formulation given by \eqref{Schrodlim1.1}-\eqref{Schrodlim1.3}, we can, as above, treat $\theta_A=\theta_A(t,x)$ and $\theta_B=\theta_B(t,x,y)$
as independent variables provided that we choose initial data for $\theta_B$ satisfying $\Ah\theta_B|_{t=0} = 0$.

These limit equations define our notion of a classical-quantum limit. The classical nature
of the limit system is clear from equation \eqref{Schrodlim1.2}, which is of the form of a Hamilton-Jacobi equation from classical mechanics. We note that it only depends on the configuration coordinate $x$ of the first system, which allows
us to interpret the first system as being classical and $x$ as a classical configuration coordinate. Further, we observe that the first equation \eqref{Schrodlim1.1} is the conservation law for the
probability density
\eqn{rhodef}{
\rho = R^2
}
that is common to both quantum and classical systems. This can be seen by introducing the velocity field
\eqn{vfield}{
\vv = (v_1,v_2) := \biggl(\frac{1}{m_1}\del{x}\theta_A , \frac{1}{m_2}\del{y}\theta_B\biggr),
}
which allows us to write \eqref{Schrodlim1.1} as the conservation law
\leqn{claw}{
\del{t}\rho + \Div (\vv \rho) = 0.
}
It is also apparent that equation \eqref{Schrodlim1.3} is closely related to the quantum mechanical equation
satisfied by the phase function with the main differences being the appearance of the complementary projection operator $\Bh$ and the velocity
coupling term $-\frac{1}{m_1}\del{x}\theta_A \del{x}\theta_B= -v_x\del{x}\theta_B$. Without these terms, \eqref{Schrodlim1.3} would be identical to the quantum
mechanical equation for the phase function, i.e. a Hamilton-Jacobi equation with a quantum potential. As we show below in Section \ref{loclim}, by using Lagrangian coordinates
adapted to the velocity field $v_x$ along with a change of variables, equations \eqref{Schrodlim1.1} and \eqref{Schrodlim1.3}
can be cast in the form of an exact Schr\"{o}dinger equation, albeit with a modified time-dependant potential. This justifies the interpretations of the second system
as being quantum mechanical.

An important point regarding these limit equations is that we can expect that solutions to these equations to approximate solutions to order $\ep$ of the fully quantum system
on time intervals $[0,T]$ for which the solutions to the limit equations remain smooth.
This is because the limit equations are derived from taking the $\ep \searrow 0$
of the $\ep$-dependent 2-particle Schr\"{o}dinger equation \eqref{Schrod}, and for related limits, such as the classical limit,
there exists many rigorous mathematical results of this character available in the literature; for example, see \cite{Bardos_et_al:2015} and reference
therein.
Thus, we expect that if $(R^\ep,\theta_A^\ep, \theta^\ep_B)$ is a solution of
the full quantum mechanical system, then
\eqn{psiapprox}{
R^\ep =  R + \Ord{\ep}{}, \quad  \theta_A^\ep = \theta_A + \Ord{\ep}{}, \AND \theta_B^\ep = \theta_B + \Ord{\ep}{}
}
on some fixed time interval [0,T] where $(R,\theta_A,\theta_B)$ is a solution of the limit equations \eqref{Schrodlim1.1}-\eqref{Schrodlim1.3}.
In this way, a solution $(R,\theta_A,\theta_B)$ of the limit equations
can be interpreted as the zeroth order expansion term in an $\ep$ expansion of a quantum mechanical solution of
the $\ep$-dependent 2-particle Schrodinger equation \eqref{Schrod}. From this starting point, higher order expansion
terms in $\ep$ can be derived in a standard fashion. We derive the equations satisfied by the first order correction in Section \ref{app2}.

From similar results for the classical limit (see \cite{Bardos_et_al:2015}), we can expect that time interval $[0,T]$ on which solutions
to the limit equations will approximate the fully quantum system will be determined by the time interval for which the limit equations
admit smooth solutions. As we establish below in Section \ref{loclim}, this time interval is determined by the time of existence for
the classical Hamiltonian flow associated to the Hamilton-Jacobi equation \eqref{Schrodlim1.2} and the time of existence of smooth solutions to \eqref{Schrodlim1.2}, which is
well known to be governed by the formation of caustics. Presumably, in line with known results for the classical limit, see again \cite{Bardos_et_al:2015}, it should be possible
using a more complicated multi-phase construction to extend the time of the validity of the approximation to arbitrary time intervals. We will not consider this aspect further
here and instead leave it for future consideration.

\subsect{noback}{No quantum backreaction}
In the setting of quantum-classical hybrid schemes, the notion of quantum backreaction refers to the effect
of the quantum subsystem on the classical one. Here, we show that the classical-quantum limit equations
derived above do not allow for quantum backreaction. This
is in concordance with the theoretical studies carried out in \cite{CaroSalcedo:1999,Sahoo:2003,Salcedo:1996,Salcedo:2012}. It also makes complete
sense from the origin of the limit equations as an extreme mass ratio limit. In such a limit, one does not expect the light particles to
affect the heavy ones at the ``zeroth order''. However, it is worth noting
that for general choices of the potential $V=V(x,y)$, the classical equations do have an effect on the quantum ones, and so, in
general, there is a non-trivial
interaction between the two systems.

It is important to note
that it is still possible to augment our limit equations with interactions that do allow
for a quantum backreaction on the classical subsystem. Indeed, as it will be shown in Section \ref{app1} by adding
appropriate interaction terms to the limit equations, we can obtain the Hall-Reginatto hybrid
scheme equations which do allow for a quantum backreaction \cite{Hall:2008,HallReginatto:2005,ReginattoHall:2009}. Of course,
this means leaving the realm of quantum mechanics and limits thereof into the regime of classical-quantum
hybrids, which are theories distinct from quantum mechanics as we know it. On the other hand,
a quantum backreaction of the quantum system  on the classical one can be obtained within the framework of quantum mechanics by
considering the first order correction to the limit equations that are derived in Section \ref{app2}. From our
point of view, this is a more fundamental and promising direction for understanding the ``true'' interaction
between nearly classical systems and fully quantum ones.

In order to establish that our classical-quantum limit equations do not generate a quantum backreaction,
we consider the marginal density
\eqn{rho1def}{
\rho_1(t,x) = \int_{\Rbb} \rho(t,x,y)\, dy,
}
which contains all of the statistical information relating to the classical system. It follows from integrating \eqref{claw} together with the fact that $v_1$ is independent of $y$ that $\rho_1$ satisfies
\eqn{claw1}{
\del{t}\rho_1 + \del{x}(v_1 \rho_1) = 0.
}
Noting that this equation and \eqref{Schrodlim1.2} form a closed system of evolution equations for
$(\theta_A,\rho_1)$, it follows immediately that the quantum subsystem does not have any effect on the classical system and establishes
the absence of a quantum backreaction.

\subsect{loclim}{A local formulation}
Due to the appearance of the averaging operator in the limit equations, they are manifestly nonlocal, and consequently, difficult to interpret, and from a computational point of view, are more expensive to solve compared to local systems. Fortunately, solutions to the
limit equations can be obtained from solutions of the following local system:
\lalin{Schrodlim2}{
\del{t}R &= -\frac{1}{m_1}\del{x}\theta_A \del{x}R -
 \frac{1}{2m_1}\del{x}^2\theta_A R -\frac{1}{m_2}\del{y}\thetat_B \del{y}R - \frac{1}{2 m_2}\del{y}^2\thetat_B R, \label{Schrodlim2.1}\\
\del{t}\theta_A &= -\frac{1}{2 m_1}\bigl(\del{x}\theta_A\bigr)^2
-U, \label{Schrodlim2.2} \\
\del{t}\thetat_B &= -\frac{1}{m_1}
\del{x}\theta_A\del{x}\thetat_B
-\frac{1}{2 m_2}\bigr(\del{y}\thetat_B\bigr)^2
 + \frac{1}{2 m_2} \frac{\del{y}^2 R}{R} - V. \label{Schrodlim2.3}
}
Indeed, given a solution
\eqn{equivA}{
\bigl(\theta_A=\theta_A(t,x), \thetat_B=\thetat_B(t,x,y), R=R(t,x,y)\bigr)
}
of \eqref{Schrodlim2.1}-\eqref{Schrodlim2.3},
a straightforward computation using the relations \eqref{Bh2.1}-\eqref{Bh3}
establishes that
\leqn{equivB}{
\bigl(\theta_A,\theta_B := \Bh\thetat_B, R\bigr)
}
is a solution of \eqref{Schrodlim1.1}-\eqref{Schrodlim1.3}.

We can further simplify the local system \eqref{Schrodlim2.1}-\eqref{Schrodlim2.3} by introducing a Lagrangian coordinate for the classical subsystem
that is adapted to the vector field
\leqn{v1Lag}{
v_1(t,x) = \frac{1}{m_1} \del{x}\theta_A(t,x),
}
which determines the motion of the classical particle.
The Lagrangian transformation is defined using the flow $F(t,x)$ of $v_1(t,x)$, which is uniquely determined by
the initial value problem
\lalin{v1Flow}{
\del{t}F(t,x) &= v_1(t,F(t,x)), \label{v1Flow.1} \\
F(0,x) &= x.  \label{v1Flow.2}
}
We note that the initial condition \eqref{v1Flow.2} implies that
\eqn{Flowidata}{
\del{x}F(0,x) = 1,
}
and this, in turn, guarantees that
the condition
\leqn{jac2}{
\del{x}F(t,x) > 0
}
holds on some time interval $[0,T]$. The condition \eqref{jac2} is precisely the necessary condition for the Lagrangian coordinate transformation to be invertible; for the remainder of this section, we will assume that
\eqref{jac2} is satisfied.

Defining
\lalin{alphavars}{
\alpha_A(t,x) &:= \theta_A(t,F(t,x)), \label{alphavars.1}\\
\alpha_B(t,x,y) &:= \thetat_B(t,F(t,x),y) \label{alphavars.2}
\intertext{and}
\Rc(t,x,y) &:= R(t,F(t,x),y)\sqrt{\del{x}F(t,x)}, \label{alphavars.3}
}
it follows directly from \eqref{v1Lag}, \eqref{v1Flow.1} and the chain rule that
\lalin{dtalpha}{
\del{t}\alpha_A(t,x) &= \del{t}\theta_A(t,F(t,x)) + \frac{1}{m_1}\bigl(\del{x}\theta_A(t,F(t,x))\bigr)^2,
\label{dtalpha.1}\\
\del{t}\alpha_B(t,x,y) &= \del{t}\thetat_B(t,F(t,x),y) + \frac{1}{m_1}
\del{x}\thetat_B(t,F(t,x),y)\del{x}\theta_A(t,F(t,x)) \label{dtalpha.2}
\intertext{and}
\del{t}\Rc(t,x,y) &= \biggl[\del{t}\Rc(t,F(t,x),y)+\frac{1}{m_1}\del{x}\Rc(t,F(t,x),y)\del{x}\theta_A(t,F(t,x))
\notag \\
&\hspace{2.5cm} +\frac{1}{2m_1}\del{t}R(t,F(t,x),y)\del{x}^2\theta_A(t,F(t,x))\biggr]\sqrt{\del{x}F(t,x)}. \label{dtalpha.3}
}
Given a solution
\eqn{sollimA}{
\bigl(\theta_A=\theta_A(t,x), \thetat_B=\thetat_B(t,x,y), R=R(t,x,y)\bigr)
}
of \eqref{Schrodlim2.1}-\eqref{Schrodlim2.3}, we see from \eqref{dtalpha.1}-\eqref{dtalpha.3}
that the pair
\eqn{sollimB}{
\bigl(\alpha_B=\alpha_B(t,x,y),\Rc=\Rc(t,x,y)\bigr)
}
satisfies the equations
\alin{prSchrod}{
\del{t} \Rc = -\frac{1}{m_2}\del{y}\alpha_B\del{y}\Rc -\frac{1}{m_2}\del{y}^2\alpha_B \Rc,\\
\del{t}\alpha_B = -\frac{1}{m_2}\bigl(\del{y}\alpha_B\bigr)^2 + \frac{1}{2m_2}\frac{\del{y}^2\Rc}{\Rc}
-\Vt
}
where
\eqn{Vtdef}{
\Vt(t,x,y) := V(F(t,x),y),
}
which we recognize as the polar representation of Schr\"{o}dinger's equation. This establishes
that the wave function $\psit$ defined by
\leqn{psitdef}{
\psit(t,x,y) := e^{i\alpha_B(t,x,y)}\Rc(t,x,y)
}
satisfies the time-dependent Schr\"{o}dinger equation
\leqn{tdSchrod}{
i\del{t}\psit = -\frac{1}{2m_2}\del{y}^2 \psit + \Vt.
}

The above equation shows that the $y$ variable is best interpreted as the configuration variable of a 1-particle quantum
subsystem in which the motion of the classical particle appears as a time dependent parameter in the
potential $\Vt(t,x,y)=V(F(t,x),y)$. If $V$ is chosen independent of $x$, then $\Vt=V(y)$ and
\eqref{tdSchrod} reduces
to precisely a 1-particle Schr\"{o}dinger equation.

We can further simplify the equation \eqref{tdSchrod} by replacing the flow $F(t,x)$ with the Hamiltonian flow associated with the
Hamilton-Jacobi equation \eqref{Schrodlim2.2}. To see this, consider the Hamiltonian
\eqn{cHam}{
H_1(x,p) = \frac{1}{2m_1}p^2 + U(x)
}
and let
\eqn{HflowA}{
\Fc(t,x,p) = (X(t,x,p),P(t,x,p))
}
denote the correponding Hamiltonian flow, i.e. $\Fc$ satisfies
\lalin{HflowB}{
\del{t}X(t,x,p) &= \frac{\del{}H_1}{\del{}p}\bigl(\Fc(t,x,p)\bigr), \label{HflowB.1} \\
\del{t}P(t,x,p) &= -\frac{\del{}H_1}{\del{}x}\bigl(\Fc(t,x,p)\bigr), \label{HflowB.2} \\
\Fc(0,x,p) &= (x,p). \label{HflowB.3}
}
Then from the method of characteristics, see \cite[\S 3.2 \& 3.3]{Evans:2010}, it follows that
\leqn{HflowC}{
F(t,x) = X\bigl(t,x,\del{x}\theta_A(0,x)\bigr),
}
and that the solution $\theta_B$ to the Hamilton-Jacobi equation \eqref{Schrodlim2.2} is
given in terms of the Hamiltonian flow by the formula
\leqn{HflowD}{
\theta_A(t,x) - \theta_A(0,G(t,x)) + \int_0^t\biggl(\frac{1}{2m_1}\bigl|P\bigl(\tau,\xt,\del{x}\theta_A(0,\xt)\bigr)\bigr|^2-U\bigl(F(\tau,\xt)\bigr)\biggr)\, d \tau \Bigl|_{\xt=G(t,x)}
}
where $G(t,x)$ is for constant $t$ the inverse of $F(t,x)$, i.e. $G(t,F(t,x))=x$.

From the above calculations, it is clear
that solutions to the classical-quantum limit system can be generated from solutions to the time-dependent 1-particle Schr\"{o}dinger equation
\eqref{tdSchrod} and the classical Hamiltonian flow equations \eqref{HflowB.1}-\eqref{HflowB.3} via the transformations
\eqref{equivB}, \eqref{alphavars.2}, \eqref{alphavars.3}, \eqref{psitdef}, and \eqref{HflowC}, and the formula \eqref{HflowD}. In this way, we can view equations
\eqref{tdSchrod} and \eqref{HflowB.1}-\eqref{HflowB.3} as an equivalent formulation of the classical-quantum limit equations at least on time intervals for which $F(t,x)$
is invertible, that is before the onset of caustics. Moreover, since \eqref{tdSchrod} is linear, it is the time of existence for the classical Hamiltonian
flow and the formation time for caustics that limit our ability to solve the classical-quantum limit equations.

\sect{app1}{The origin of nonlocal signalling in the Hall-Reginatto hybrid scheme}

To understand the origin of the nonlocal signalling in the Hall-Reginatto hybrid scheme,
we consider the case where the classical-quantum limit equations \eqref{Schrodlim0.1}-\eqref{Schrodlim0.2} arise from
the limit of two commuting 1-particle Hamiltonians by assuming that the potential $V$ is a function of
$y$ only, that is
\eqn{Vset}{
V = V(y).
}
We then write the classical-quantum limit equations as
\leqn{CQeqn}{
\del{t}\begin{pmatrix} \rho \\ \theta \end{pmatrix} = X_{C}(\rho,\theta) + X_{Q}(R,\theta)
}
where
\leqn{CQv1}{
X_{C}(\rho,\theta) = \begin{pmatrix} -\del{x}\biggl(\frac{1}{m_1}(\del{x}\Ah \theta)\rho \biggr) \\
 -\frac{1}{2 m_1}\Bigl[\bigl(\del{x}\Ah\theta\bigr)^2 + 2\del{x}\Ah\theta\del{x}\Bh\theta\Bigr]
-U
\end{pmatrix}
}
and
\leqn{CQv2}{
X_{Q}(\rho,\theta) = \begin{pmatrix} -\del{y}\biggl(\frac{1}{m_2}(\del{y} \theta)\rho \biggr)\\
  \Bh\biggl[
  -\frac{1}{2 m_2}\bigr(\del{y}\theta\bigr)^2
 + \frac{1}{2 m_2} \frac{\del{y}^2 \sqrt{\rho}}{\sqrt{\rho}}\biggl] - V
 \end{pmatrix}.
}
Since equation \eqref{CQeqn} arises from the limit of a 2-particle quantum system whose evolution
is governed by the sum of the two commuting 1-particle Hamiltonian operators
\eqn{1partHam}{
\Hh_1 = -\frac{\ep}{2m_1}\del{x}^2 + \frac{1}{\ep}U \AND \Hh_2 =
-\frac{1}{2 m_2}\del{y}^2 + V,
}
and
the $\ep\searrow 0$ limit of the 1-particle Hamiltonian operators $\Hh_1$ and $\Hh_2$, in our variables, are given by
the two vector fields \eqref{CQv1} and \eqref{CQv2}, respectively, it follows that the flows of these vector fields
must commute, and moreover, from the discussion above, that $X_C$  and $X_Q$ generate the evolution of independent classical and quantum subsystems, respectively. Thus, if we let $F^{X_{C}}_t$, $F^{X_{Q}}_t$, and $F^{X_{C}+X_{Q}}_t$ denote the flows of
the vector fields $X_{C}$, $X_{Q}$, and $X_{C}+X_{Q}$, respectively, then
$F^{X_{C}}_t$ and $F^{X_{Q}}_t$ must commute, i.e.
\leqn{flowcom}{
F_t^{X_{C}}\circ F_s^{X_{Q}} = F_s^{X_{Q}}\circ F_t^{X_{C}},
}
and $F^{X_{C}+X_{Q}}_t$ is given by
\leqn{flowdecomp}{
F^{X_{C}+X_{Q}}_t = F_t^{X_{C}}\circ F_t^{X_{Q}},
}
which allows us to write solutions of \eqref{CQeqn} as
\eqn{solrep}{
\begin{pmatrix}R(t)\\ \theta(t) \end{pmatrix} = F_t^{X_{C}}\circ F_t^{X_{Q}}(R(0),\theta(0)).
}
The two properties \eqref{flowcom} and \eqref{flowdecomp} reflect nothing more than the independence
of the evolution of the classical and quantum subsystems.

Next, since the quantum system satisfies the no non-local signalling
property with respect to position measurements of the 1-particle subsystems by virtue of the evolution being generate by
the sum $\Hh_1 + \Hh_2$ of commuting Hamiltonian operators, it follows automatically that the limit system
must also satisfy this property since the position measurements are represented by $\ep$ independent maps in our formalism.
In contrast to our equations \eqref{CQeqn} for non-interacting classical-quantum systems,  those
of Hall and Renginatto are given by \cite{HallReginatto:2005}
\leqn{HReqn}{
\del{t}\begin{pmatrix}\rho \\ \theta \end{pmatrix} = X_{HR}(\rho,\theta)
}
where
\eqn{HRv}{
X_{HR}(\rho,\theta) = \begin{pmatrix} -\del{x}\biggl(\frac{1}{m_1}(\del{x} \theta) \rho \biggr)-\del{y}\biggl(\frac{1}{m_2}(\del{y} \theta) \rho \biggr) \\
  -\frac{1}{2 m_1}\bigr(\del{x}\theta\bigr)^2
  -\frac{1}{2 m_2}\bigr(\del{y}\theta\bigr)^2
 + \frac{1}{2 m_2} \frac{\del{y}^2 \sqrt{\rho}}{\sqrt{\rho}} - (U+V)
 \end{pmatrix}.
}
These equations are known \cite{Hall_et_al:2012} to suffer from a non-local signalling effect with respect to position measurements.

Writing the Hall-Reginatto equations as
\eqn{HReqnA}{
\del{t}\begin{pmatrix}\rho \\ \theta \end{pmatrix} = X_{C}(\rho,\theta)+X_{Q}(\rho,\theta) + X_{I}(\rho,\theta)
}
where
\eqn{HReqnA}{
X_{I}(\rho,\theta) :=  X_{HR}-X_{C}-X_Q = \begin{pmatrix} -\del{x}\biggl(\frac{1}{m_1}\del{x}( \Bh\theta )\rho\biggr)  \\
-\frac{1}{2 m_1}\bigr(\del{x} \Bh \theta\bigr)^2
 +\Ah\Bigl[ -\frac{1}{2 m_2}\bigr(\del{y}\theta\bigr)^2
 + \frac{1}{2 m_2} \frac{\del{y}^2 \sqrt{\rho}}{\sqrt{\rho}} \Bigr] \end{pmatrix},
}
we see that $X_{I}$ represents a non-local interaction term that
alters the evolution generated
by the vector field $X_C + X_Q$.
From this, we conclude that the non-local interaction term $X_{I}$ is responsible for the
non-local signaling effect in the Hall-Reginatto equations \eqref{HReqn}.
It is also clear from the above calculations that we can
obtain any of Hall-Reginatto classical-quantum hybrid equations
from the articles \cite{Hall:2008,HallReginatto:2005,ReginattoHall:2009} by adding a suitable interaction term to $X_C+X_Q$.

\sect{app2}{The first order correction}

As discussed above, the classical-quantum limit equations derived in Section \ref{limiteqns} may be viewed
as the zeroth order equations in an $\ep$-expansion. For the purpose of improving the approximation and
also determining the backreaction of the second subsystem on the first, it is necessary to include
the effects of higher order corrections. In this section, we derive the equations satisfied by the first
order correction.

We begin the derivation by decomposing $R^\ep$, $\theta^\ep_A$, and $\theta^\ep_B$ as
\leqn{Rthetaexp}{
R^\ep= R + \ep \mu^\ep, \quad \theta^\ep_A = \theta_A + \ep \nu^\ep \AND \theta^\ep_B = \theta_B+\ep \omega^\ep
}
where $\nu^\ep$ and $\omega^\ep$ satisfy $\Ah \nu^\ep =\nu^\ep$ and $\Ah \omega^\ep = 0$, respectively, and, as above, $(R,\theta_A,\theta_B)$ satisfy the classical-quantum limit equations \eqref{Schrodlim1.1}-\eqref{Schrodlim1.3}.
Substituting \eqref{Rthetaexp} into the equations \eqref{Schrod2a.1}-\eqref{Schrod2a.3}, we
find, with the help of \eqref{Schrodlim1.1}-\eqref{Schrodlim1.3}, that the triple
$(\mu^\ep, \nu^\ep,\omega^\ep_B)$ satisfies the following equations:
\lalin{Schrod3}{
\del{t}\mu^\ep &= -\frac{1}{m_1}\Bigl(\del{x}\theta_A \del{x}\mu^\ep + \bigl(\del{x}\nu^\ep +
\del{x}\theta_B+\ep \del{x}\omega^\ep\bigr)(\del{x}R+ \ep\del{x}\mu^\ep\bigr)\Bigr)
-\frac{1}{2 m_1}
\Bigl(\del{x}^2\theta_A \mu^\ep  \notag\\
& \hspace{0.4cm}+ \bigl(\del{x}^2 \nu^\ep +\del{x}^2\theta_B+\ep \del{x}^2\omega^\ep\bigr)
\bigl(R+\ep \mu^\ep\bigr)\Bigr) - \frac{1}{m_2}(\del{y}\theta_B+\ep \del{y}\omega^\ep) \del{y}\mu^\ep
-\frac{1}{2 m_2} (\del{y}^2\theta_B+\ep \del{y}^2\omega^\ep) \mu^\ep, \label{Schrod3.1} \\
\del{t}\nu^\ep & = -\frac{1}{2m_1}\Bigl(2 \del{x}\theta_A \del{x}\nu^\ep + \ep \bigl( (\del{x}\nu^\ep)^2+
\Ah \bigl(\del{x}\theta_B+\ep \del{x}\omega^\ep\bigr)^2 \bigr)\Bigr)
\notag \\
&\hspace{1.0cm} + \Ah\biggl[\frac{\ep}{2 m_1}\biggl(\frac{\del{x}^2R+\ep \del{x}^2 \mu}{R+
\ep\mu^\ep}\biggr)
  -\frac{1}{2 m_2} \bigl(\del{y}\theta_B+\ep \del{y}\omega^\ep\bigr)^2 + \frac{1}{2 m_2}\biggl(\frac{\del{y}^2R+\ep \del{y}^2 \mu^\ep}{R+\ep\mu^\ep}\biggr)\biggr],  \label{Schrod3.2} \\
\del{t} \omega^\ep &= \Bh \biggl[-\frac{1}{2m_1}\Bigl(2\del{x}\theta_A\del{x}\omega^\ep + 2\del{x}\nu\del{x}\theta_B + 2\ep \del{x}\omega^\ep\del{x}\nu
+ \bigl(\del{x}\theta_B+\ep \del{x}\omega^\ep)^2\Bigr)
 +\frac{1}{2m_1}\biggl(\frac{\del{x}^2R+\ep \del{x}^2 \mu^\ep}{R+\ep\mu^\ep}\biggr) \notag \\
 &\hspace{1.0cm}
  -\frac{1}{ 2 m_2}\bigl(2\del{y}\theta_B\del{y}\omega^\ep + \ep(\del{y}\omega^\ep)^2\bigr) + \frac{1}{2m_2}\frac{1}{\ep}\biggl[ \biggl(\frac{\del{y}^2R+\ep \del{y}^2 \mu^\ep}{R+\ep\mu^\ep}-\frac{\del{y}^2R}{R}\biggr) \biggr]
 \biggr]. \label{Schrod3.3}
}
We observe that the equations \eqref{Schrodlim1.1}-\eqref{Schrodlim1.3} and \eqref{Schrod3.1}-\eqref{Schrod3.3} for the variables $(R,\theta_A,\theta_B,\mu^\ep,\nu^\ep,\omega^\ep)$ are equivalent to the Schr\"{o}dinger equation
\eqref{Schrod} via the transformation
\eqn{SchrodtransB}{
\psi = e^{i\bigl(\frac{\theta_A}{\ep} + \nu^\ep + \theta_B+\ep \omega^\ep\bigr)}(R+\ep \mu^\ep).
}

Taking the formal $\ep \searrow 0$ limit of the equations \eqref{Schrod3.1}-\eqref{Schrod3.3}, we obtain the equations
\lalin{Schrodlim3}{
\del{t}\mu &= -\frac{1}{m_1}\Bigl(\del{x}\theta_A \del{x}\mu + \bigl(\del{x}\nu + \del{x}\theta_B)\del{x}R\Bigr) \notag \\
& \hspace{2.0cm} -\frac{1}{2 m_1}\Bigl(\del{x}^2\theta_A \mu + \bigl(\del{x}^2 \nu +\del{x}^2 \theta_B\bigr)R\Bigr) - \frac{1}{m_2}\del{y}\theta_B \del{y}\mu -\frac{1}{2 m_2}\del{y}^2\theta_B \mu, \label{Schrodlim3.1} \\
\del{t}\nu & = -\frac{1}{m_1}\del{x}\theta_A \del{x}\nu + \Ah\biggl[
  -\frac{1}{2 m_2} (\del{y}\theta_B)^2 + \frac{1}{2 m_2}\frac{\del{y}^2R}{R}\biggr],  \label{Schrodlim3.2} \\
\del{t}\omega &= \Bh \biggl[-\frac{1}{2m_1}\Bigl(2\del{x}\theta_A\del{x}\omega + 2\del{x}\nu\del{x}\theta_B +
 \bigl(\del{x}\theta_B\bigr)^2\Bigr) \notag\\
&\hspace{2.0cm} +\frac{1}{2m_1}\frac{\del{x}^2R}{R}
  -\frac{1}{ m_2}\del{y}\theta_B\del{y}\omega + \frac{1}{2m_2}\biggl(\frac{\del{y}^2\mu}{R}-\frac{\mu \del{y}^2 R}{R^2}\biggr) \biggr]. \label{Schrodlim3.3}
}
These equations define the first order correction to the limit equations, and we can expect, if $(R^\ep,\theta^\ep_A,\theta^\ep_B)$ is a solution of
the full quantum mechanical system, then
\eqn{psifirstexp}{
R^\ep =  R + \ep \mu + \Ord{\ep}{2}, \quad \theta^\ep_A = \theta_A + \ep \nu + \Ord{\ep}{2} \AND \theta^\ep_B = \theta_B + \ep\omega + \Ord{\ep}{2}
}
on some fixed time interval [0,T] where $(R,\theta_A,\theta_B)$ solves \eqref{Schrodlim1.1}-\eqref{Schrodlim1.3} and
$(\mu,\nu,\omega)$ solves \eqref{Schrodlim3.1}-\eqref{Schrodlim3.3}. Higher order corrections can be determined in the usual manner.

We emphasize that the system consisting of the equations \eqref{Schrodlim1.1}-\eqref{Schrodlim1.3} and \eqref{Schrodlim3.1}-\eqref{Schrodlim3.3} represents a fully consistent first order
approximation to the Schr\"{o}dinger equation that should be accurate for modeling the interaction between
particles of disparate masses in the regime where the large mass particles are effectively classical.
We expect these equations should be of use in computational chemistry and also the theory
of measurement where interactions between quantum and classical systems play a distinguished role.

\sect{disc}{Discussion and outlook}

In this article, we have introduced a new approach to analyzing the interaction between classical and quantum
systems that is based on taking the classical-quantum limit of multi-particle Schr\"{o}dinger equations. The
classical-quantum limit equations obtained by this procedure describe the interaction between classical and quantum systems
and possess many desirable properties that are inherited from the multi-particle quantum systems.

We considered two applications of the classical-quantum limit equations with the first being to identify
the source of the non-local signalling that is present in the ``non-interacting'' classical-quantum hybrid equations of Hall and Reginatto. More specifically, we showed that the non-local signalling in the Hall-Reginatto scheme
is due to a non-local correction to our classical-quantum limit equations, which we contend should be taken
as the ``true'' form for the equations that govern non-interacting classical-quantum systems in the Hall-Reginatto hybrid scheme.

In the second application, we viewed the classical-quantum limit equations as the equations
satisfied by the zeroth order expansion in $\ep$ of a fully quantum solution. We then, in the standard fashion, derived
the system of equations satisfied by the first order correction. Together, the limit equations and their
first order correction determine a fully consistent first order
approximation to the multi-particle Schr\"{o}dinger equation. We expect that these equations
will be useful for modeling the interaction between
particles of disparate masses in the regime where the large mass particles are effectively classical.

The results derived in this article can be taken in many directions. From the point of view of concrete applications,
the most important is to develop analytical and numerical techniques to solve
the limit equations \eqref{Schrodlim1.1}-\eqref{Schrodlim1.3} along with the first order
corrections \eqref{Schrodlim3.1}-\eqref{Schrodlim3.3} in order to determine their
suitability for modelling systems that consist of nearly classical subsystems
interacting with fully quantum ones. Important examples include molecular reactions and the interaction of measuring
devices with quantum systems. We are currently working on this and hope to
have progress to report on in the near future.

\begin{acknowledgments}
This work was partially supported by the Australian Research Council grant FT1210045. I thank the referee for their comments
and criticisms, which have served to improve the content
and exposition of this article.
\end{acknowledgments}

\appendix

\sect{crep}{Equivalence of the classical-quantum limits}

As discussed in Remark \ref{Arem}, there is a classical-quantum limit associated to each operator satisfying \eqref{Ah2}-\eqref{Ah4}.
In this appendix, we show that these classical-quantum limits are equivalent by showing that there is an invertible transformation
between the limits associated to different operators. An immediate consequence of this is that all choices of the operator
$\Ah$ lead to physically equivalent classical-quantum limits.

\subsect{trans}{Transformation formula}
To establish the transformation formula and its invertibility, we suppose that $\Ah$ and $\Ahu$ are two operators that satisfy the properties \eqref{Ah2}-\eqref{Ah4},
and we let $\Bh = \id - \Ah$ and $\Bhu =\id -\Ahu$ denote the complementary projection operators. Following the prescription \eqref{thetadef}, we
define
\lalin{trans1}{
\theta^\ep := \ep \Ah \Theta^\ep + \Bh \Theta^\ep, \notag \\ 
\thetau^\ep := \ep \Ahu \Theta^\ep + \Bh \Theta^\ep, \notag 
}
where $\Theta^\ep$ and $R^\ep$ are as defined previously by the polar decomposition \eqref{Rthetadef}. These
two definitions lead to two classical-quantum limits defined by
\lalin{trans2}{
(\theta,R) := \lim_{\ep\searrow 0} (\theta^\ep,R^\ep),  \label{trans2.1}\\
(\thetau,R) := \lim_{\ep\searrow 0} (\thetau^\ep,R^\ep). \label{trans2.2}
}

From the relation \eqref{thetadef1}, we observe that
\leqn{trans3}{
\frac{1}{\ep}\Ah\theta^\ep + \Bh\theta^\ep = \Theta^\ep = \frac{1}{\ep}\Ahu\, \thetau^\ep + \Bhu\, \thetau^\ep.
}
Applying $\Ahu$ to this expression yields
\leqn{trans4}{
\frac{1}{\ep}\Ahu\Ah\theta^\ep + \bigl(\Ahu-\Ahu\Ah\bigr)\theta^\ep = \frac{1}{\ep}\Ahu\, \thetau^\ep,
}
where in deriving this we used the property $\Ahu^2=\Ahu$. By assumption, $\Ah$ satisfies $\del{y}\Ah =0$, and so,
for any function $\xi=\xi(x,y)$, $\Ah\xi =\phi$ for some function $\phi=\phi(x)$. But since $\Ahu(\phi)=\phi$ for all functions $\phi=\phi(x)$ by assumption,
it follows that
\leqn{trans5}{
\Ahu\Ah= \Ah,
}
and from similar considerations, that
\leqn{trans6}{
\Ah\Ahu= \Ahu.
}
Using \eqref{trans5}, we obtain from \eqref{trans4} that
\leqn{trans7}{
\Ah\theta^\ep + \ep \bigl(\Ahu-\Ah\bigr)\theta^\ep = \Ahu\, \thetau^\ep.
}
Taking the limit $\ep\searrow 0$ gives
\leqn{trans8}{
\Ahu\,\thetau  = \Ah \theta.
}
Substituting \eqref{trans7} into \eqref{trans3} gives
\eqn{trans9}{
\Bhu\,\thetau^\ep = \Bh\theta^\ep +\bigl(\Ah-\Ahu\bigr)\theta^\ep,
}
which, in turn, yields
\eqn{trans9}{
\Bhu\,\thetau = \Bh\theta +\bigl(\Ah-\Ahu\bigr)\theta.
}
by taking the limit $\ep\searrow 0$.
Adding this to \eqref{trans8} then yields the relation
\leqn{trans10}{
\thetau = \Th\theta
}
where the operator $\Th$ is defined by
\leqn{trans11}{
\Th = \id +\Ah-\Ahu.
}

The operator $\Th$ is invertible with inverse given by
\leqn{trans12}{
\Th^{-1} = \id + \Ahu-\Ah.
}
To see this, we observe that
\lalin{trans13}{
\Th^{-1}\Th &= (\id +\Ahu-\Ah)(\id +\Ah-\Ahu), \notag \\
&= \id +\Ah-\Ahu + \Ahu-\Ah + \Ahu\Ah-\Ahu^2-\Ah^2+\Ah\Ahu, \notag \\
&=\id,  \label{trans13.1}
}
where in deriving the last equality we used the identities $\Ah^2=\Ah$, $\Ahu^2=\Ahu$, $\Ahu\Ah=\Ah$ and $\Ah\Ahu=\Ahu$.
By a similar calculation, we also find that
\leqn{trans14}{
\Th\Th^{-1} = \id.
}
Together, \eqref{trans13.1} and \eqref{trans14} show that $\Th$ is invertible with inverse given by \eqref{trans12}.

Equation \eqref{trans10} with $\Th$ given by \eqref{trans11} defines the relationship between the two  classical-quantum limits
\eqref{trans2.1} and \eqref{trans2.2}, and establishes the existence of an invertible transformation relating the two limits. As discussed above, this establishes the physical equivalence of the limits.

\subsect{crepex}{Examples of other operators satisfying \eqref{Ah2}-\eqref{Ah4}}
There are many operators $\Ah$ that satisfy the properties \eqref{Ah2}-\eqref{Ah4} beyond the those defined by
\eqref{Ah1}. A more general class is defined by
\leqn{Aclass1}{
\Ah(\phi)(x,y) = \int \alpha(z)\phi(x,z)\, dz,
}
where $\alpha(y)$ is any function satisfying
\leqn{Aclass2}{
\int\alpha(y)\, dy = 1.
}
We observe that we can recover the operators defined by \eqref{Ah1} by setting
\leqn{Aclass3}{
\alpha(y) = \begin{cases} (b-a)^{-1} & \text{if $y\in (a,b)$} \\
\hspace{0.4cm} 0 & \text{otherwise} \end{cases}.
}
It is also clear that by choosing functions $\alpha(y)$ satisfying \eqref{Aclass2}
that are distinct from the step-functions \eqref{Aclass3} will yield operators that
are not of the type \eqref{Ah1}, but nonetheless satisfy properties \eqref{Ah2}-\eqref{Ah4}.

Another class of operators distinct from \eqref{Aclass1} is defined by
\leqn{Aclass4}{
\Ah(\phi)(x,y) = \phi(x,a),
}
where, here, $a\in \Rbb$ is an arbitrary, but fixed constant. From the Fundamental Theorem of Calculus, which holds for regular enough functions $\phi$, we see that
\eqn{Aclass5}{
\phi(x,a) = \frac{d\;}{dy} \int_{a}^y \phi(x,z)\, dz\biggl|_{y=a} = \lim_{b\searrow a} \frac{1}{b-a}\Bigl( \int_{a}^b \phi(x,z)\, dz - \int_{a}^a \phi(x,z)\, dz\Bigr)
= \lim_{b\searrow a} \frac{1}{b-a}\int_{a}^b \phi(x,z)\, dz.
}
From this it is clear that the class \eqref{Aclass4} of operators is a limiting case of the operators defined by \eqref{Ah1} or equivalently \eqref{Aclass1} with $\alpha$ defined
by \eqref{Aclass3}.

It is worth noting at this point that although all operators $\Ah$ satisfying the properties \eqref{Ah2}-\eqref{Ah4} are equivalent from the point of view of formal limits, the mathematical properties of the operators can be quite different. For example, the operators $\Ah$ defined
using averaging procedure \eqref{Aclass1} are well-defined, assuming appropriate integrability properties for the function $\alpha$, on functions
that are much rougher compared to those on which the operators defined via \eqref{Aclass4} can act. Considerations like this will likely become
important in any attempt to go beyond formal considerations and rigorously establish the existence of the classical-quantum limit introduced in this article.
We expect that only particular choices of the operators $\Ah$ will be suitable for rigorous arguments.

\bibliographystyle{amsplain}
\bibliography{hybrid}

\providecommand{\bysame}{\leavevmode\hbox to3em{\hrulefill}\thinspace}
\providecommand{\MR}{\relax\ifhmode\unskip\space\fi MR }
\providecommand{\MRhref}[2]{%
  \href{http://www.ams.org/mathscinet-getitem?mr=#1}{#2}
}
\providecommand{\href}[2]{#2}
\begin{thebibliography}{10}

\bibitem{Agostini_et_al:2007}
F.~Agostini, S.~Caprara, and G.~Ciccotti, \emph{Do we have a consistent
  non-adiabatic quantum-classical mechanics?}, EPL \textbf{78} (2007), 30001.

\bibitem{Anderson:1995}
A.~Anderson, \emph{Quantum backreaction on ``classical'' variables}, Phys. Rev.
  Lett. \textbf{74} (1995), 621--625.

\bibitem{Bardos_et_al:2015}
C.~Bardos, F.~Golse, P.~Markowich, and T.~Paul, \emph{On the classical limit of
  the {S}chr\"{o}dinger equation}, Discrete Cont. Dynam. Systems \textbf{35}
  (2015), 5689--5709.

\bibitem{BoucherTraschen:1988}
W.~Boucher and J.~Traschen, \emph{Semiclassical physics and quantum
  fluctuations}, Phys. Rev. D \textbf{37} (1988), 3522--3532.

\bibitem{Bousquet_et_al:2011}
David Bousquet, Keith~H. Hughes, David~A. Micha, and Irene Burghardt,
  \emph{Extended hydrodynamic approach to quantum-classical nonequilibrium
  evolution. {I}. {T}heory}, J. Chem. Phys. \textbf{134} (2011), 064116.

\bibitem{CaroSalcedo:1999}
J.~Caro and L.L. Salcedo, \emph{Impediments to mixing classical and quantum
  dynamics}, Phys. Rev. A \textbf{60} (1999), 842--852.

\bibitem{Diosi_et_al}
L.~Di\'osi, N.~Gisin, and W.T. Strunz, \emph{Quantum approach to coupling
  classical and quantum dynamics}, Phys. Rev. A \textbf{61} (2000), 022108.

\bibitem{DiosiHalliwell:1998}
L.~Di\'osi and J.J. Halliwell, \emph{Coupling classical and quantum variables
  using continuous quantum measurement theory}, Phys. Rev. Lett. \textbf{81}
  (1998), 2846--2849.

\bibitem{Elze:2012b}
H-T Elze, \emph{Four questions for quantum-classical hybrid theory}, J. Phys.:
  Conf. Ser. \textbf{361} (2012), 012004.

\bibitem{Elze:2012a}
\bysame, \emph{Linear dynamics of quantum-classical hybrids}, Phys. Rev. A
  \textbf{85} (2012), 052109.

\bibitem{Elze:2013}
\bysame, \emph{Quantum-classical hybrid dynamics - a summary}, J. Phys.: Conf.
  Ser. \textbf{442} (2013), 012007.

\bibitem{Evans:2010}
L.C. Evans, \emph{Partial differential equations}, $2^{\text{nd}}$ ed., AMS,
  2010.

\bibitem{Fratino_et_al:2014}
L.~Fratino, A.~Lampo, and H-T Elze, \emph{Entanglement dynamics in a
  quantum–classical hybrid of two q-bits and one oscillator}, Physica Scripta
  \textbf{2014} (2014), 014005.

\bibitem{Hall:2008}
M.J.W. Hall, \emph{Consistent classical and quantum mixed dynamics}, Phys. Rev.
  A \textbf{78} (2008), 042104.

\bibitem{HallReginatto:2005}
M.J.W. Hall and M.~Reginatto, \emph{Interacting classical and quantum
  ensembles}, Phys. Rev. A \textbf{72} (2005), 062109.

\bibitem{Hall_et_al:2012}
M.J.W. Hall, M.~Reginatto, and C.M. Savage, \emph{Nonlocal signaling in the
  configuration space model of quantum-classical interactions}, Phys. Rev. A
  \textbf{86} (2012), 054101.

\bibitem{Horenko_et_al:2004}
I.~Horenko, M.~Weiser, B.~Schmidt, and C.~Sch\"{u}tte, \emph{Fully adaptive
  propagation of the quantum-classical {L}iouville equation}, 2004,
  pp.~8913--8923.

\bibitem{Kapral:1999}
R.~Kapral and G.~Ciccotti, \emph{Mixed quantum-classical dynamics}, The Journal
  of Chemical Physics \textbf{110} (1999), 8919--8929.

\bibitem{Kisil:2005}
V.V. Kisil, \emph{A quantum-classical bracket from p -mechanics}, EPL
  \textbf{72} (2005), 873.

\bibitem{Lampo_et_al:2014}
A.~Lampo, L.~Fratino, and H-T Elze, \emph{Mirror-induced decoherence in hybrid
  quantum-classical theory}, Phys. Rev. A \textbf{90} (2014), 042120.

\bibitem{Meier:1993}
C.~Meier, \emph{Mixed quantum-classical treatment of vibrational decoherence},
  Phys. Rev. Lett. \textbf{93} (2004), 173003.

\bibitem{PeresTerno:2001}
A.~Peres and D.R. Terno, \emph{Hybrid classical-quantum dynamics}, Phys. Rev. A
  \textbf{63} (2001), 022101.

\bibitem{PrezhdoBrooksby:2001}
O.V. Prezhdo and C.~Brooksby, \emph{Quantum backreaction through the {B}ohmian
  particle}, Phys. Rev. Lett. \textbf{86} (2001), 3215--3219.

\bibitem{PrezhdoKisil:1997}
O.V. Prezhdo and V.V. Kisil, \emph{Mixing quantum and classical mechanics},
  Phys. Rev. A \textbf{56} (1997), 162--175.

\bibitem{ReginattoHall:2009}
M.~Reginatto and M.J.W. Hall, \emph{Quantum-classical interactions and
  measurement: a consistent description using statistical ensembles on
  configuration space}, Journal of Physics: Conference Series \textbf{174}
  (2009), 012038.

\bibitem{Sahoo:2003}
D.~Sahoo, \emph{Mixing quantum and classical mechanics and uniqueness of
  {P}lanck's constant}, Journal of Physics A: Mathematical and General
  \textbf{37} (2004), 997.

\bibitem{Salcedo:1996}
L.L. Salcedo, \emph{Absence of classical and quantum mixing}, Phys. Rev. A
  \textbf{54} (1996), 3657--3660.

\bibitem{Salcedo:2012}
\bysame, \emph{Statistical consistency of quantum-classical hybrids}, Phys.
  Rev. A \textbf{85} (2012), 022127.

\end{thebibliography}


\end{document}